# Epitaxial growth of high-quality GaAs on Si(001) using ultrathin buffer layers

Kun Cheng[1,2,3], Tianyi Tang[1,2,3], Wenkang Zhan[1,2], Zhenyu Sun[1,2], Bo Xu[1,2], Chao Zhao[1,2,a], and Zhanguo Wang[1,2]

**AFFILIATION**

[1]Key Laboratory of Semiconductor Materials Science, Institute of Semiconductors, Chinese Academy of Sciences & Beijing Key Laboratory of Low Dimensional Semiconductor Materials and Devices, Beijing 100083, China.

[2]College of Materials Science and Opto-Electronic Technology, University of Chinese Academy of Sciences, Beijing 101804, China.

[3]Contributed equally to this work.

[a] Author to whom correspondence should be addressed: zhaochao@semi.ac.cn.

**ABSTRACT**

The direct growth of III-V semiconductors on silicon holds tremendous potential for photonics applications. However, the inherent differences in their properties lead to defects in the epitaxial layer, including threading dislocations (TDs), antiphase boundaries (APBs), and thermal cracks, which significantly impact devices performance. Current processes struggle to suppress these defects simultaneously, necessitating the development of methods to inhibit TDs and APBs in a thin buffer on silicon. In this study, a GaSb buffer layer is introduced during GaAs epitaxy on a silicon (001) substrate. This approach successfully suppresses defect formation by promoting the formation of interfacial misfit dislocation (IMF) arrays at both the AlSb/Si and GaAs/GaSb interfaces. The resulting GaAs layer exhibits a step-flow surface with a rough mean square of approximately 3.8 nm and a full width at half maximum of 158 arcsec. Remarkably, the growth is achieved without any observable interfacial intermixing. Building on this platform, InAs/GaAs quantum dots (QDs) are grown with a density of $3.8 \times 10^{10}$ cm$^{-2}$, emitting at a wavelength of 1288 nm. This breakthrough holds immense promise for the development of high-quality GaAs films with reduced defect densities on silicon for O band lasers, laying the foundation for the mass production of silicon-based integrated circuits.

The monolithic integration of III-V lasers on silicon has garnered considerable attention as a promising solution for on-chip light sources in silicon-based photonic integrated circuits due to its advantages, including cost-effectiveness, compatibility with complementary metal-oxide-semiconductor (CMOS) fabrication processes, compact device sizes, and high integration capabilities.[1,2] Nonetheless, the direct growth of III-V compounds on silicon poses substantial challenges due to lattice mismatch, thermal mismatch, and polarity mismatch. These challenges lead to the formation of defects such as antiphase domains (APDs), TDs, and thermal cracks within the epitaxial layers, ultimately compromising the quality of the epitaxial films.[3-6] Researchers have utilized silicon substrates with specific miscut angles from the (001) direction to suppress APDs.[7] However, this approach is not CMOS- compatible.[8] Another approach involves pre-patterning on-axis silicon substrates to achieve APD-free GaAs buffer layers.[9] Nonetheless, this method requires complex micro-processing and specialized substrate treatment. High-quality GaAs buffer layers were also attained by regulating the aluminum content within the AlGaAs seed layer.[10] To diminish TD, superlattice (SL) layers and thermal cycle annealing were applied.[11,12] However, pursuing high-quality GaAs necessitates a thick buffer layer, increasing the risk of thermal cracks during cooling. These thermal cracks significantly impact the lifetime and yield of laser devices, impeding the realization of silicon-based photonics integrated circuits. The current epitaxy processes cannot simultaneously suppress the APDs, TDs and thermal cracks. Therefore, there is a need to develop a new technique to inhibit the TDs and APDs in a thin buffer on CMOS-compatible Si.

Antimonides alleviate strains by generating arrays of interfacial misfit (IMF) dislocations rather than forming 60° TDs.[13] This behavior is attributed to the significant lattice mismatch (12%) between antimonides and silicon, making avoiding thick buffer layers in the growth processes possible. Notably, the lattice mismatch between GaAs and GaSb is approximately 7.8%, indicating the feasibility of over-growth of a high-quality GaAs layer with strain relieved through IMF dislocation arrays.[14] However, there is an essential concern that GaSb is susceptible to reacting with $As_2$, which poses challenges for producing larger areas of GaAs.[15,16] Therefore, careful consideration must be given to the growth process to ensure the production of high-quality GaAs layers.

In this letter, we report the achievement of direct growth of high-quality GaAs buffer layers on on-axis silicon. Our innovative methodology integrates a thin GaSb buffer layer and leverages IMF dislocations to suppress TD while employing a self-annihilation mechanism to mitigate APD.[17,18] Simultaneously, our approach is designed to prevent the undesirable $As_2$ influx into the GaSb layer. Subsequently, InAs/GaAs quantum dots (QDs) are grown on the high-quality GaAs buffer layer for emission at 1.3 μm. We employ a range of advanced characterizations to gain comprehensive insights into the mechanisms governing film growth and crystallographic structures. Atomic force microscopy (AFM) is utilized for surface morphology analysis, while transmission electron microscopy (TEM) and X-ray diffractometry (XRD) provide detailed assessments of crystallographic structures. Additionally, photoluminescence (PL) spectroscopy is employed to evaluate the QDs' luminescent properties. These meticulous characterizations enable us to thoroughly understand the growth processes and the exceptional quality of the resulting materials. All samples are grown using molecular beam epitaxy (MBE, Compact 21 T). Before loading into the buffer chamber, the Si substrate was pre-treated by soaking in hydrofluoric acid solution to remove the surface oxide, rinsed in deionized water, and then dried by high-purity nitrogen. Before transferring into the growth chamber, the Si substrate was outgassed at 400 °C in the buffer chamber. Then the substrate was treated by annealing at 1000 °C for 40 min to desorb the hydrogen atoms terminating on the Si surface. The (2 × 2) surface reconstruction pattern was observed by reflection high-energy electron diffraction (RHEED), indicating a pure Si(001) surface. The 5 nm AlSb nucleation layer was grown on Si substrate at 430 °C, followed by a GaSb layer grown at the same temperature. Subsequently, a GaSb layer with a thickness of 1045 nm was grown at 490 °C. With the temperature reduced to 335 °C in the $Sb_2$ environment, the Sb source was turned off to avoid the reaction between $As_2$ and the grown GaSb layer. Meanwhile, the As and Ga sources were turned on simultaneously, beginning the GaAs growth with a thickness of 200 nm. Finally, another layer of GaAs with a thickness of 300 nm was deposited with the same growth rate at 550 °C to complete the GaAs layer and the whole thickness reached 1.55 μm. Then The InAs/GaAs QDs were grown on the GaAs layer as mentioned-above and it consists of InGaAs/GaAs superlattice

(SL) layers, n-type GaAs, n-type AlGaAs, InAs QDs layer, p-type AlGaAs, p-type GaAs cap layer, InAs QDs surface layer.

The surface morphologies were determined by AFM with an MFP 3D scanning probe microscopy system in the atmosphere. The crystallographic structures were studied by XRD using a Bruker D8 Discover Plus diffractometer. The interface properties were evaluated by TEM with a high-resolution TEM system (HRTEM, FEI Talos 200S) operating at 200 kV. The elemental mapping was characterized using a scanning transmission electron microscope equipped with energy-dispersive x-ray spectroscopy (STEM-EDX) with an atomic resolution analytical electron microscopy system (Thermo Fisher scientific Helios5) operating at 200 kV. The TEM and STEM-EDX observation samples were prepared by focused ion beam (FIB) milling.

The AFM image in FIG. 3(a) reveals a surface of the GaAs buffer with no visible antiphase boundaries (APBs) emerging. The surface has a relatively smooth profile, measuring a root mean square (RMS) roughness of 3.8 nm. It is important to note that this roughness value is slightly higher than that of the GaSb buffer (2.1 nm) in FIG. 1(b), hinting at the potential presence of defects, such as microtwins, which may have formed at the interface and extended to the surface. The circular and rectangular dark points observed on the surface likely indicate planar defects originating from stacking faults[19,20]. It is hypothesized that the presence of stacking faults in our sample results from the 60° dislocation dissociating into two Shockley partial dislocations[21]. Although there are no visible APBs on the surface, this does not guarantee the complete absence of APBs throughout the entire structure. It suggests that while APBs might exist, they did not reach the sample surface.[22] Additionally, the observed RMS roughness value is well within the acceptable ranges for subsequent laser structure processing, suggesting compatibility with the downstream device fabrication processes.

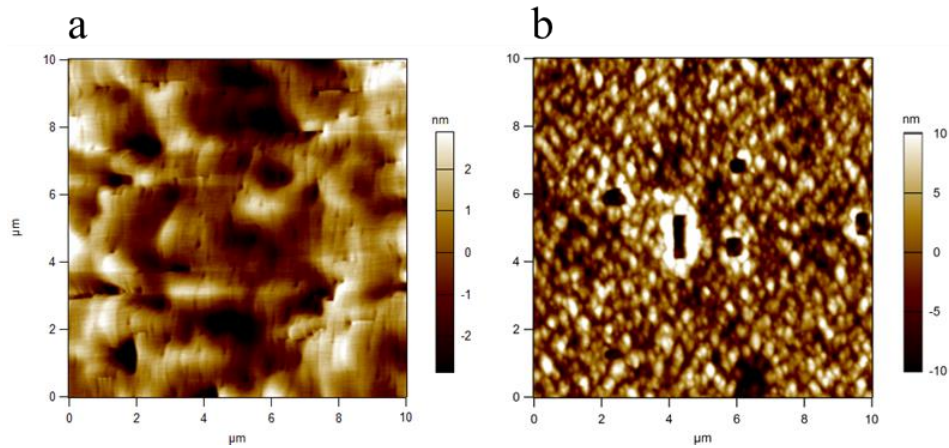

FIG. 1. 10 μm × 10 μm AFM topographic images of (a) GaSb buffer and (b) GaAs buffer.

The elemental distribution at the interface of the samples was examined using STEM-EDX. FIG. 2(a)-(d) show the EDX elemental mapping of the structure consisting of GaAs/GaSb/Si from top to bottom, with $As_2$ soaking 30 s on the GaSb surface. The GaSb surface is rather rough due to the interaction between $As_2$ and GaSb, forming nano pits and interfacial mixing. Subsequent GaAs growth coalesces over these nano pits, resulting in nanovoids and amorphous defects, thereby deteriorating the material quality, and inducing a significant density of TDs in the subsequent layer. Contrastingly, FIG. 2(e)-(h) show the EDX elemental mapping of the structure consisting of GaAs/GaSb from top to bottom and display a distinct scenario where the Ga and As sources were simultaneously turned on. The elements distribution of the interface indicates no element intermixing between the GaAs/GaSb interface. The interface appears notably smoother and devoid of nano pits, highlighting that simultaneous turning on Ga and As sources minimizes the strong reaction between $As_2$ and GaSb. At low temperature, GaAs nucleates on the GaSb surface with a uniform size, followed by their connection. Subsequent growth of a high-temperature GaAs layer promotes the annihilation of defects. Consequently, the interface remains free of undesirable nano pits, presenting a smooth and favorable interface structure.

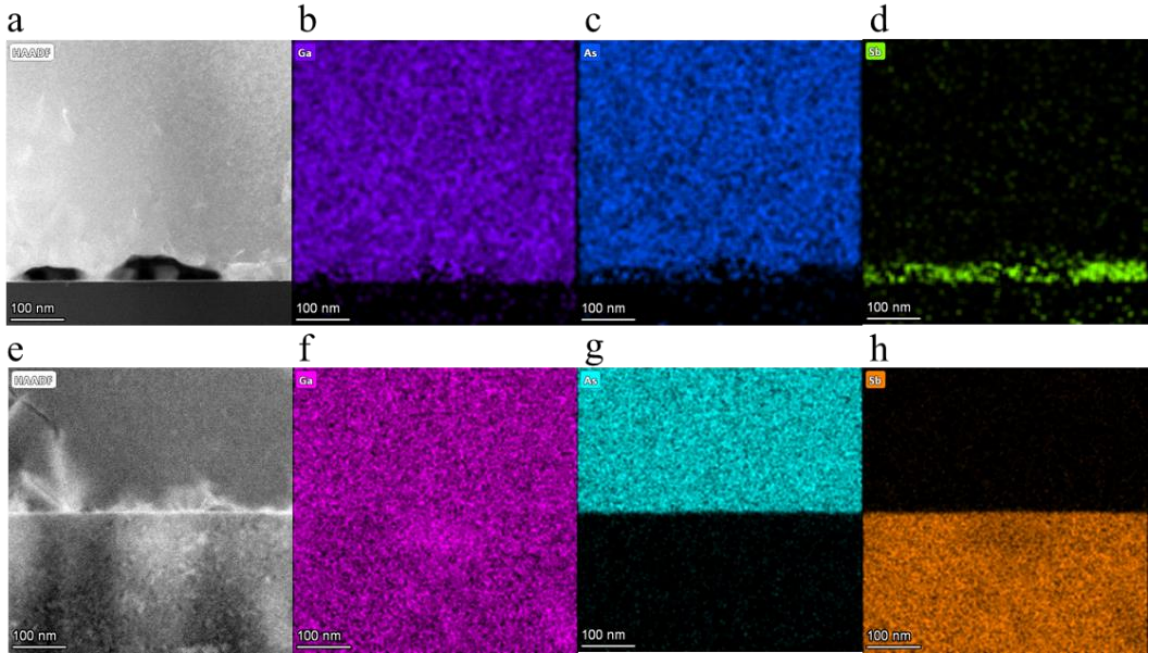

FIG. 2. STEM-HADDF image (a) and EDX elemental mapping of Ga, As, and Sb, respectively, (b)-(d), of GaAs/GaSb interface with $As_2$ soaking. The lower interface is GaSb/Si interface, upper interface is GaAs/GaSb interface. STEM-HADDF image (e) and EDX elemental mapping of Ga, As, and Sb, respectively, (f)-(h), of GaAs/GaSb interface without $As_2$ soaking. The interface is GaAs/GaSb interface.

The XRD rocking curves of the (004) planes of the GaSb buffer layer and GaAs layer are depicted in FIG. 3(a), (b). Notably, GaAs exhibits a lower full width at half maximum (FWHM) of 158 arcsecs compared to GaSb (230 arcsecs), potentially due to the formation of an IMF dislocation array. The TD density of GaAs is approximately $9 \times 10^7$ cm$^{-2}$ from the FWHM, according to Bhatnagar[23]. Additionally, FIG. 3(c) and (d) display the sample's reciprocal space mapping (RSM). The mapping reveals that the diffraction peaks of GaAs, GaSb, and Si align with the (004) direction, indicating the desired orientation of the GaAs epitaxial layer without inclination. The vertical ($a_\perp$) and parallel ($a_\parallel$) lattice constants of the GaAs layer were calculated using the reciprocal lattice vectors $q_x$ and $q_z$ of the GaAs peak in the (224) RSM.[24] The calculated results are presented in Table I. Remarkably, the calculated results, together with the diffraction peak, align closely with the (224) direction shown in FIG. 3(d), demonstrating that the GaAs layers have achieved a nearly 100% of relaxation and excellent crystalline quality.

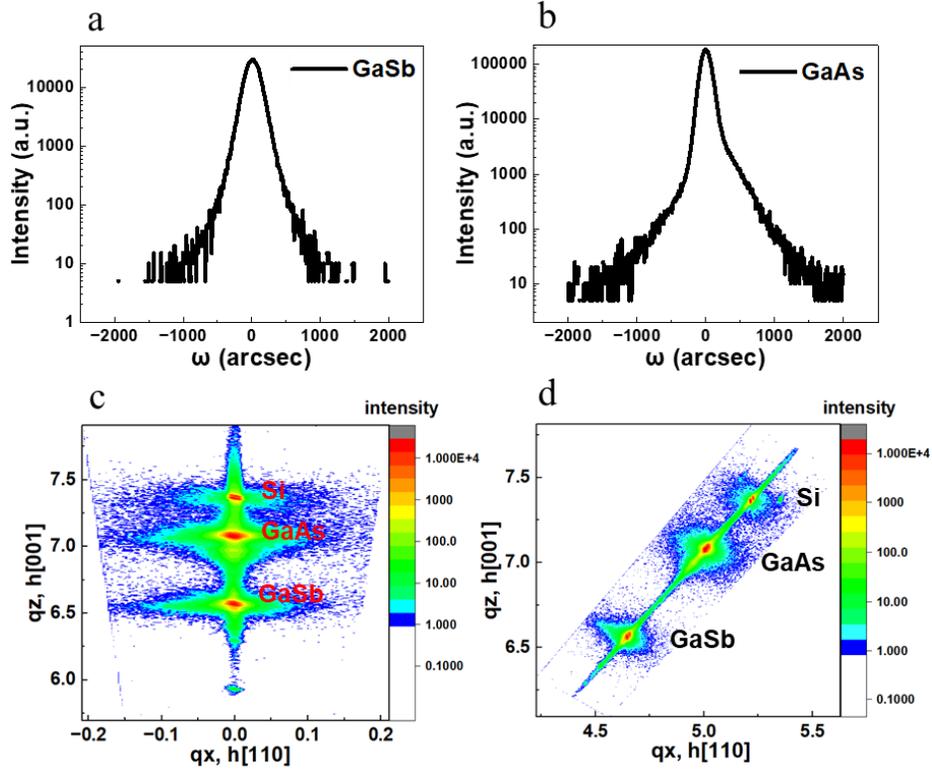

FIG. 3. The XRD rocking curve of the (004) planes of (a) GaSb buffer layer and (b) GaAs layer. X-ray RSMs of (c) (004) and (d) (224) of GaAs grown on Si(001) with GaSb buffer.

Table I. Lattice constants, relaxation, FWHM from (004) ω-scan of GaAs and GaSb

| Structure | $a_\perp$ | $a_\parallel$ | Relaxation | FWHM |
| --- | --- | --- | --- | --- |
| GaAs | 5.658Å | 5.657Å | 99.0% | 158 |
| GaSb | 6.098Å | 6.102Å | 99.9% | 230 |

TEM was employed to study the microstructure of the GaAs/GaSb interface, shown in FIG. 4(a), revealing an IMF dislocation array near the interface. The bright and dark bands at the interface represent the bond-bending around the IMF dislocations.[25] In particular, the dark bands represent the IMF dislocation array itself.[26] It is assumed that the IMF dislocation array fully compensates for the lattice mismatch between the GaAs and GaSb layers. Based on this assumption, we can calculate the separation (S) between adjacent misfit dislocations in the uniform IMF dislocation array using equation (1). This equation incorporates the length of the Burgers vector b of a misfit dislocation in the layer and the lattice mismatch f between the substrate and the epitaxy layer, providing valuable insights into the interface's structural characteristics.[27]

$$S = \frac{b}{f} \qquad (1)$$

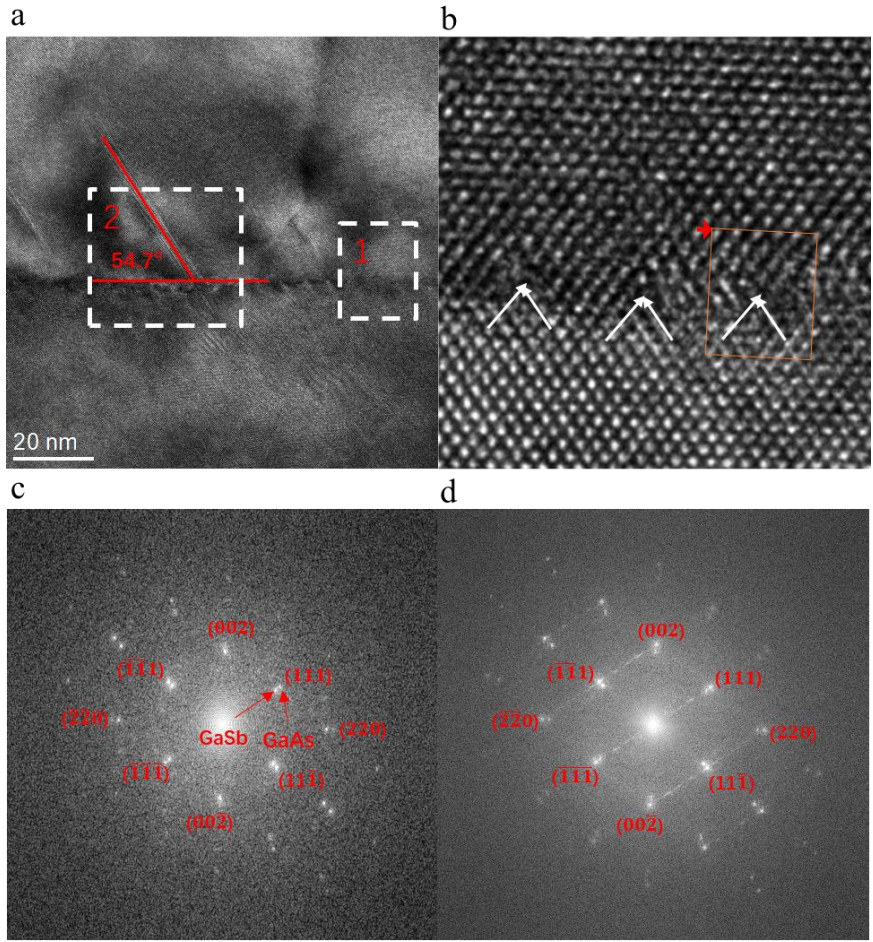

FIG. 4. (a) TEM image of GaAs/GaSb interface. (b) HRTEM image of GaAs/GaSb interface. FFT images of (c) region 1 and (d) region 2.

The separation of two misfit dislocations observed in FIG. 4(a) is approximately 5.52 nm, which agrees well with the calculated value of 5.53 nm. This value corresponds precisely to 14 GaAs lattice sites grown over 13 GaSb lattice sites along the [110] directions. A more in-depth investigation of the GaAs/GaSb interfacial crystal structure, as shown in FIG. 4(b), reveals the presence of 90° dislocations, indicated by white arrows. These dislocations follow the right-handed start-to-finish (RHSF) convention,[21] with Burgers vectors of $a/2[110]$, classifying them as 90° dislocations. They are concentrated near the interface, with their formation potentially resulting from the reaction between two 60° dislocations residing on intersecting (111) glide planes. The additional half-planes of these 60° dislocations, corresponding to (111) and (11$\bar{1}$) glide planes, run parallel to the white arrows highlighted in FIG. 4(b). This dislocation reaction is energetically favorable and can be represented as[21]:

$$\frac{a}{2}[101] + \frac{a}{2}[01\bar{1}] = \frac{a}{2}[110] \tag{2}$$

To further analyze the structural characteristics, A fast Fourier transform (FFT) was carried out to analyze the structural characteristics further, and the images are provided in FIG. 4(c) and (d), with marked diffraction spots.

In region 1, distinct from the GaAs/GaSb interface, we observe uniform and ordered crystal lattices free of defects, supported by the presence of two diffraction spots corresponding to GaAs and GaSb in the FFT pattern in FIG. 4(c), indicated by two arrows. The FFT pattern confirms the epitaxial growth of GaAs thin films on the GaSb buffer layer. To assess the lattice characteristics, we calculated the lattice spacing of GaAs using TEM images in FIG. 4(a) and the diffraction spot spacing of GaAs in the FFT pattern in FIG. 4(b). The calculated values ($d_{GaAs(111)}$, $d_{GaAs(220)}$, and $d_{GaAs(002)}$) are presented in Table II, along with their theoretical values. Remarkably, each GaAs lattice spacing closely matches its theoretical value, providing that the lattice-mismatched strain in the GaAs thin film has been effectively and nearly completely relieved. This underscores the high-quality epitaxial growth and structural integrity achieved in the GaAs thin film on the GaSb buffer layer.

Table II. Evaluation of $d_{GaAs(111)}$, $d_{GaAs(220)}$, and $d_{GaAs(002)}$ based on the HRTEM image in FIG. 4(b), and the FFT pattern in FIG. 4(c), together with the theoretical values.

| | $d_{GaAs(111)}$ (nm) | $d_{GaAs(220)}$ (nm) | $d_{GaAs(002)}$ (nm) |
| --- | --- | --- | --- |
| HRTEM image | 0.333 | 0.202 | 0.286 |
| FFT pattern | 0.322 | 0.195 | 0.278 |
| Theoretical value | 0.326 | 0.200 | 0.283 |

In region 2, there's a unique feature called twinned GaAs, characterized by a different atomic arrangement compared to epitaxial GaAs in region 1. These microtwins are oriented along the (111) plane with a 54.7° inclination from the GaSb buffer layer. As shown in FIG. 4(c), symmetric additional diffraction spots around the (111) and ($\bar{1}\bar{1}\bar{1}$) planes indicate the presence of microtwins. Microtwins can introduce lattice orientation inclination, leading to these extra diffraction spots, confirming their existence.[28] Microtwins often form during the growth of III-V compounds on (111) surfaces, and the steps on the buffer layer likely contribute to their formation, consistent with prior research.[29] According to Machida et al.,[30] when grown on (111) surface, adatoms can occasionally undergo a 60° rotation, forming a wurtzite structure as a stacking fault on the (111) surface. Subsequently, twins with a different atomic arrangement compared to the buffer layer are generated in zinc blende structures. Adatoms can rotate by another 60° on the (111) surface, returning to the original lattice structure, which provides an alternative explanation for the formation of GaAs microtwins, as it offers a gliding surface for 60° threading dislocations[31], contributing to the degradation of the surface morphology in GaAs.

Several mechanisms have been proposed to explain the selectivity of 60° or 90° dislocation formation in lattice mismatch heteroepitaxy. The misorientation of the substrate from (001) leads to a misfit dislocation (MD) split. In contrast, the stress accumulated in the epitaxy layer leads to the change of dislocation type.[32,33] According to Huang et al.,[34] when GaSb is grown on GaAs, it exhibits an IMF array growth mode. Thus, the growth mode of GaAs on top of GaSb is predicted to follow similar principles as observed in previous studies.[35] Both theoretical and experimental investigations concerning IMF dislocation arrays suggest that the pure 90° edge dislocations form due to periodic disruptions in chemical bonds, which agrees well with the growth conditions associated with the IMF array growth mode.[13,27] The GaAs islands formed on the GaSb buffer contribute to the formation of the IMF dislocation array, thus affecting the lattice strain-relaxation of GaAs. The strain between GaAs and GaSb contributes to the TDs bending in the interface, prompting IMF dislocations arrays connecting, which is helpful to the epitaxial growth of GaAs on GaSb buffer grown on Si(100) substrate. Table III summarizes the results of the structural properties of GaAs/Si. Our work demonstrates a balance between reducing thermal cracks and TD density, even though the TD density remains high in the GaAs buffer layer. This method suggests the potential for minimizing APBs and TDs while preventing thermal cracks in a thin buffer. Based on this platform, InAs/GaAs QDs grown on Si is feasible.

Table III. Summary of literature data about the GaAs/Si films.

| Ref. | Method | Substrate | Buffer Thickness(μm) | TD Density(cm$^{-2}$) |
|---|---|---|---|---|
| 9 | MBE+MOCVD | On-axis Si | 2.02 | $7\times10^7$ |
| 10 | MBE | On-axis Si | 2.3 | $1.1\times10^8$ |
| 36 | MBE | On-axis Si | 1 | $8\times10^8$ |
| 37 | MBE | On-axis GaP/Si | 3.3 | $7.2\times10^6$ |
| 38 | MBE | On-axis GaP/Si | 2.55 | $2\times10^6$ |
| 39 | MOCVD | 0.5° offcut Si | 1.65 | $2.7\times10^7$ |
| This work | MBE | On-axis Si | 1.55 | $9\times10^7$ |

The InAs/GaAs QDs are grown on the obtained GaAs virtual substrate and characterized by AFM and PL. As shown in FIG. 5(a), InAs QDs are evenly distributed on the surface with a density of approximately $3.8\times10^{10}$ cm$^{-2}$, affirming the successful growth of InAs QDs. The room-temperature PL spectrum in FIG. 5(b) reveals a full width at half maximum (FWHM) of the ground state luminescence peak at 52.1 meV, with a corresponding peak wavelength of 1288 nm. The characterizations of InAs QDs grown on GaAs substrate are presented in FIG. 5(c) and (d), with a density of approximately $5\times10^{10}$ cm$^{-2}$ and an FWHM of the ground state luminescence peak at 43.1 meV, along with a peak wavelength of 1250 nm. Although the performance of the GaAs virtual substrate does not match that of a GaAs substrate, it still provides a viable platform for subsequent device growth.

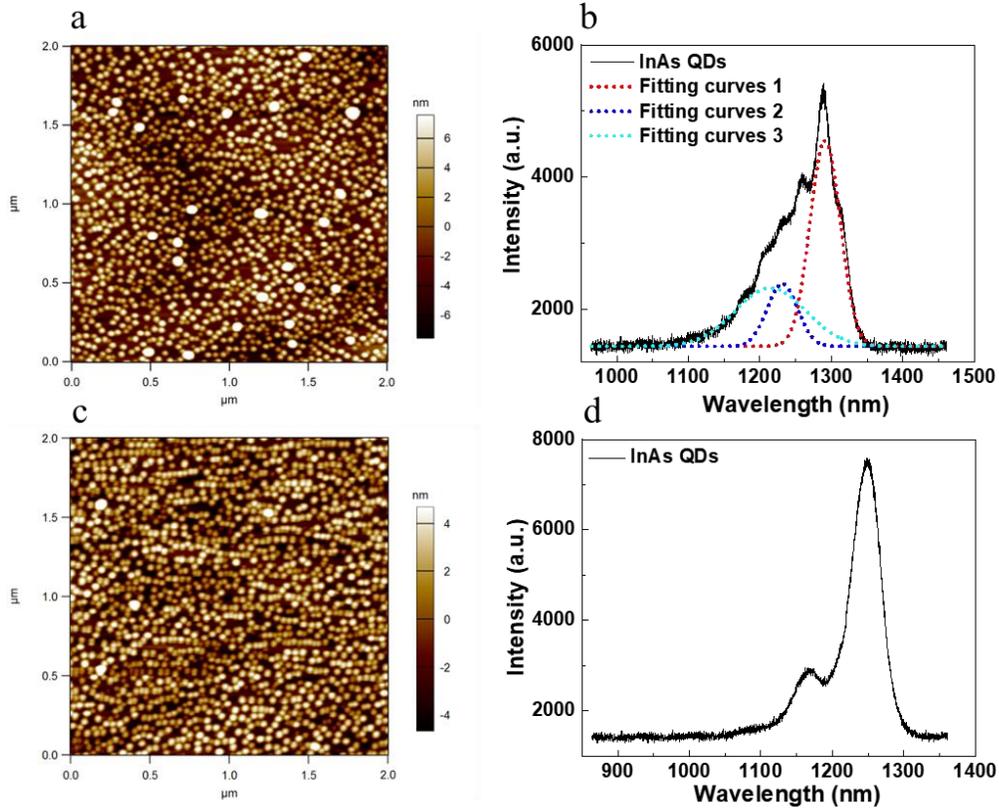

FIG. 5. (a) AFM image and (b) room-temperature PL spectrum of InAs QDs on GaAs virtual substrate. The red, blue, and cyan dot lines are fitting curves. (c) AFM image and (d) room-temperature PL spectrum of InAs QDs on GaAs substrate.

In conclusion, high-quality GaAs film and InAs/GaAs QDs are achieved by MBE direct growth on CMOS-compatible Si(001) substrate. The GaSb layer of 45 nm was grown on Si at 430°C, with an AlSb prelayer to accommodate the mismatch between GaSb and Si. GaSb layer was achieved by APD self-annihilation, and then the GaAs layer was grown by IMF growth method. The smooth GaAs/GaSb interface was demonstrated by STEM-EDX. The structure and surface morphology properties are characterized by HRTEM, XRD, and AFM, indicating that GaAs is almost relaxed on the GaSb surface by IMF dislocation array. After that, InAs/GaAs QDs were completed on the GaAs virtual substrate. This work demonstrated that a high-quality GaAs layer on Si can be realized by inserting ultrathin buffer with a thickness of hundreds of nm, thus avoiding thermal cracks. These results will be enlightening in producing high-quality GaAs thin films, with flatter and fewer crystal defects, on CMOS-compatible Si (001) substrates, opening avenues for integrating III-V QDs semiconductors on Si for high-speed, high-efficiency electronic and optoelectronic devices in the future.

## ACKNOWLEDGEMENT

This work is funded by National Key Research and Development Program of China (2021YFB2206503); National Natural Science Foundation of China (62274159); "Strategic Priority Research Program" of the Chinese Academy of Sciences (XDB43010102); CAS Project for Young Scientists in Basic Research (YSBR-056).

## AUTHOR DECLARATIONS

### Conflict of Interest Statement

The authors have no conflicts to disclose.

## AUTHOR CONTRIBUTIONS

Chao Zhao conceived the idea and designed the experiments. Tianyi Tang and Wenkang Zhan performed the MBE experiments. Tianyi Tang and Kun Cheng performed the material characterizations. Kun Cheng wrote the manuscript. Zhenyu Sun and Chao Zhao revised the manuscript. Chao Zhao and Bo Xu led the MBE effort, and Zhanguo Wang supervised the team. All authors have read, contributed to, and approved the final version of the manuscript.

## DATA AVAILABILITY

The data that support the findings of this study are available within the article.